\documentclass[11pt]{article} 

\usepackage[utf8]{inputenc} 
\usepackage[T1]{fontenc}
\usepackage{currvita}
\usepackage[noblocks]{authblk}     

\usepackage{graphicx}
\usepackage{caption}
\usepackage{subcaption}

\usepackage{geometry} 
\usepackage{booktabs} 
\usepackage{array} 
\usepackage{paralist} 
\usepackage{verbatim} 
\usepackage{subfig} 
\usepackage{url}

\usepackage{fancyhdr} 
\usepackage{sectsty}
\allsectionsfont{\sffamily\mdseries\upshape} 

\usepackage[nottoc,notlof,notlot]{tocbibind} 
\usepackage[titles,subfigure]{tocloft} 

\title{Too dynamic to fail. \\ Empirical support for an autocatalytic model of Minsky's financial instability hypothesis \\
{\small \bf The final publication is available at Springer via http://dx.doi.org/10.1007/s11403-015-0163-7}}

\author[1]{Nata\v{s}a Golo}
\affil[1]{Racah Institute for Physics, Hebrew University Jerusalem, Israel} 
\author[2]{David S. Br\'ee}
\affil[2]{Dept. of Computer Science, University of Manchester, Manchester, UK}
\author[1]{Guy Kelman}
\author[3]{Leanne Usher}
\affil[3]{Queens College, City University of New York, US}
\author[4]{Marco Lamieri}
\affil[4]{Dept. of Research, Intesa SanPaolo, Milano, Italy}
\author[1]{Sorin Solomon}

\begin{document}
 \maketitle

\begin{abstract}
Solomon and Golo \cite{Solomon 2013} have recently proposed an autocatalytic (self-reinforcing) feedback model  which   couples a  macroscopic system parameter (the interest rate),  a microscopic parameter that measures the distribution of the  states of the individual agents (the number of   firms in financial difficulty) and  a peer-to-peer network  effect (contagion across supply chain financing). 
In this model, each financial agent is characterized by its resilience to the interest rate. Above a certain  rate the interest due on the firm's financial costs exceeds its earnings and  the firm becomes susceptible to failure (ponzi). For the interest rate levels under a certain threshold level, the firm loans are smaller then its earnings and the firm becomes 'hedge.' In this paper, we fit the historical data (2002-2009) on interest rate data into our model, in order to predict the number of the  ponzi firms. We compare the prediction with the data taken from a large panel of Italian firms over a period of 9 years. We then use trade credit linkages to discuss the connection between the ponzi density and the network percolation.

We find that the 'top-down'-'bottom-up' positive feedback loop accounts for most of the Minsky crisis accelerator dynamics. 
The peer-to-peer ponzi companies contagion becomes significant only in the last stage of the crisis when the ponzi density is above a critical value.  Moreover the ponzi contagion is limited only to the companies that were not dynamic enough to substitute their distressed clients with new ones. In this respect the data support a view in which the success of the economy depends on substituting the static 'supply-network' picture with an interacting dynamic agents one. 

\end{abstract}

\section{Introduction}
\label{sec:Intro}
Firms differ in their ability to meet their financial obligations. To capture the differences, Minsky \cite{Minsky 1975}, \cite{Minsky 1975a}, \cite{Minsky 1984}, proposed three different types of financial fragility, based on the extent to which a firm is expected to be able to cover the cost of its borrowings from its  operations:  hedge financing, where both the interest rate and the amount of the loans due are covered, speculative financing, where only the interest rate is covered, and ponzi financing, where not even the interest rate is covered. This categorization is  sensitive to changes in the cost of borrowing.

Minsky believed that in the upswing of the business cycle, the central bank would increase interest rates so the proportion of speculative and ponzi firms relative to hedge firms would increase. Banks would then react by increasing  their interest rates which in turn would lead to  an increase in the number of financially fragile firms. 
Therefore, a prolonged period of economic growth would lead eventually to a crisis and a downturn.
This was known as Minsky's \cite{Minsky 1984} financial instability hypothesis.

The `Minsky moment' starts with ``a prolonged period of rapid
acceleration of debt" in which more traditional and benign borrowing is
replaced by borrowing that depends on new debt to repay existing loans.
When the `moment' occurs, then ``lenders become increasingly cautious or
restrictive, and [when] it isn't only over-leveraged structures that encounter
financing difficulties. At this juncture, the risks of systemic economic contraction
and asset depreciation become all too vivid" (definition George
Magnus, in \cite{Whalen 2008}).
If left unchecked, the Minsky moment can become a `Minsky meltdown',
spreading decline in asset values capable of producing a recession (definition by Paul McCullin, in \cite{Whalen 2008}). Even without a meltdown, the jobs market
can soften. 

A model, outlined in Section \ref{sec:AMF model}, was developed earlier and published in \cite{Solomon 2013}, which we will use to test this Minsky hypothesis. This model is based on the autocatalytic feedback loop between the number of the hedge/ponzi firms and the interest rate. To build an intuition on the autocatalytic feedback we will describe here a mathematical metaphor for the Gaia (earth) hypothesis \cite{Lovelock 1979}, the Daisyworld  model. The Daisyworld model postulates feedback loops between living species and the environmental conditions. One assumes there are two kinds of daisies: black and white. The black daisies like to live in the cold climate, because being black they absorb more energy from the sun. Thus if one starts with a cold climate, the black daisies will take over the white ones. At the planetary level having the earth surface covered by black, will retain in the earth system an increasing quantity of the solar energy reaching the earth. This will result in an increasing temperature of the earth. In the resulting warmer climate the white daisies which avoid overheating by reflecting the sun light will be advantaged.  As the white daisies start to predominate, at the planetary level, the earth will reflect part of the solar energy reaching it. The climate will cool down again and the black daisies come to flourish again. This feedback loop between the living individuals and the earth (Gaia) system as a whole are postulated by the Gaia hypothesis to bring the Gaia system, as well as its individual components, to an optimal state. The mechanism can be reinterpreted conceptually as an adaptive self-sustaining property of the Gaia system.  
Our model with firms and the interest rate,  outlined in Section ~\ref{sec:AMF model}, could be perceived as similar. In this analogy, the hedge firms could be compared with the black flowers, the interest rate with the warmth of the earth, the ponzi firms with the white flowers and the banks with the sun.

In the rest of this paper, we confront the assumptions and predictions of our model with empirical data.
We find strong support for the existence of Minsky-like “top-down”-“bottom-up” positive feedbacks.
In particular we consider:
\begin{itemize}
\item the interest rate as the global (“top”, “up”) parameter and 
\item the status (establishment, disappearance, hedge $\rightarrow$ ponzi transition) of individual firms as the individual (“bottom”, ”down”) parameters. 
\end{itemize}
We identify evidences for the Minsky scenario:
\begin{enumerate}
\item A period of optimism (2002-2005)  where the interest rate decreases and the number of companies increases.
\item A moment (2005) when interest rate starts raising and the increase in the number of companies slows down and eventually plunges.
\item A moment (2006) when many hedge companies start losing their very stable / solid status and the interest rate continues to rise.
\item The Minsky moment (2007) when there is a very sharp rise in the number of ponzi companies and a continuation of the rise in the interest rate. The Minsky accelerator is triggered: one falls into an uncontrollable feedback loop between:
\begin{itemize}
\item	the bottom process of individual companies becoming ponzi, and
\item	the top process of global credit becoming scarce and interest rate increasing.
\end{itemize}
\item In 2008 the above problems become acute enough to be addressed by the regulators:  the interest rate is exogenously cut down by the central bank in an attempt to stop the Minsky accelerator. 
\item This is however too late: in 2009 (and following years) in spite of artificially (exogenously imposed)  low interest rate the number of failures / ponzi units continues to grow and the number of companies continues to dwindle. It is the so called “liquidity trap”. 
It is only at this stage that we find peer-to-peer distress contagion signal in our data.  
The effect is limited to companies that were not able to find new customers in place of their currently financially distressed ones. This give the title of the present paper: the companies dynamic enough to break the static “supply network” concept do not fail. They find new customers and survive.  
\end{enumerate}

In the rest of this paper, we bring empirically evidence of some of the postulates of this model. We introduce the empirical data in Section~\ref{sec:Data} and the way this data was used to test the model in Section ~\ref{sec:Empirical}.

\section{A model of  Minsky's financial instability hypothesis}
\label{sec:AMF model}
A model of the propagation of innovation \cite{Cantono 2010} provided the basis for our model with  autocatalytic feedback. 
Our  model for testing the financial instability hypothesis incorporating an  autocatalytic feedback loop between the proportion of ponzi firms and the interest rate was presented previously  in \cite{Solomon 2013}. We give here just an outline.

This discrete dynamical model is a cobweb \footnote{We present a discrete iterative process where at each iteration either the quantity of loans or the interest rate adjust for loan supply costs. Following this, lowering the interest rate will decrease demand, while raising the interest rate will lower demand. This is similar to Walras groping process. The quantity of loans are accommodative and initially nominal interest rates are sticky but will adjust in accordance with the cost of lending. This is similar Marshallian process, see for instance \cite{Humphry 1992}. Our combined marshall-walras process is described in detail in \cite{Solomon 2013}. The mathematics of such iterative processes is studied in detail in the monograph \cite{Galor 2007}. For a continuous time treatment see \cite{Vercelli 2011}.} type of analysis where:
\begin{itemize}
\item  the supply of loans adjusts to meet excess demand for loans, and nominal interest rates are set after this by lenders based on their marginal cost, and 
\item the demand for loans is based on a given interest rate set by the lenders.
\end{itemize} 

Prices (interest rates) and then quantities (loans) adjust to clear the market in an iterative process. The model has two separate regimes of operation:
\begin{itemize}
\item in the period when the economy is stable and the number of risky firms is not growing, the policy of the banks on the interest rate is guided by the demand for credit and a low risk premium is expected.
\item in the period when the firms become more risky, the policy of the banks on the interest rate is guided by the number of risky firms. The higher the number of risky firms, the higher risk premium is expected, on top of the normal interest.
\end{itemize}

The model may either converge to an equilibrium, or to diverge, depending on the strength/level of the model parameters. 

A resilient firm is one that can make interest payments out of its cash flow, in contrast with a ponzi firm that must borrow to pay interest, i.e.:
\begin{equation}
income-debt * interest \enspace rate <0
\label{eq:pf}
\end{equation} 

Our resilience measure is a micro measure of firm susceptibility to macro variables, in this case the interest rate, just like the $debt/equity$ ratio. A growing firm with a lot of debt might have sufficient cash flow to pay its expenses, reinvestment, even expect to borrow more debt. The ratio is non-binding and may not indicate profitability. But such a firm is susceptible to macro interest rate rises where rising interest costs can quickly eat into profits - how quick? $Income/debt$ ratio might give an indication. Therefore, the resilience of the firms is of the key parameters of the system, and it is defined as:
\begin{equation}
resilience=\frac{income}{debt},
\label{eq:resilience}
\end{equation}
which interpretation can take on a slightly different meaning in the different regimes.

In the first regime, the firms resilience is correlated with the volume of banks loans that the firms can demand which in turn is also a function of interest rate: the higher the interest rate, the lower the volume of loans. 
We assume that this effect depends on a negative heterogeneity coefficient of a power law, so that a log plot of loan volume against interest rate will be a straight line with a negative slope. There is an interest rate which is so low that even the least robust firm can take out a loan; we assume that it is the ratio of the actual interest rate to this minimum that determines the volume of loans. These dependences are given by Eq. \ref{eq:Nloans of i}:
\begin{equation}
N_{loans}(t)=\left( \frac{i(t)}{i_{min}} \right) ^ \mu \times N_{tot}(t)
\label{eq:Nloans of i}
\end{equation}
where:
	\begin{itemize}
	\item $i_{min}$ is the minimum resilience, a constant that reflects the interest rate that the least robust firm could accept to take another loan;
	\item $N_{tot}(t)$ is the total number of firms in the data set at time $t$, and
	\item $\mu$  is the heterogeneity coefficient of firms resiliences, a negative number, being the slope of a Pareto law distribution of the resilience of the firms in the system, a cumulative probability that a particular firm will have resilience larger than a specific value, and thus be capable to take a new loan with an interest rate at this specific value.
	\end{itemize}
We expect the volume of bank loans $N_{loans}$ to be correlated with the number of companies who are capable of paying back their bank loans, i.e. the hedge companies:
\begin{equation}
N_{loans} (t+1)\sim N_{hedge}(t+1) \times \frac{N_{tot}(t+1)}{N_{tot}(t)}.
\end{equation}

In the second regime  on the other hand, we assume that  the number of firms whose resilience  is so low, at a certain moment $t$, that they are incapable of paying  even  the interest of their loans, i.e.  ponzi firms, $N_{ponzi}(t)$, is a positive function of the interest rate, $i(t)$, that is being charged for bank loans. We again assume that this relationship has a heterogeneity coefficient of a power law so that a log plot of the number of ponzi firms against the interest rate will be a straight line (but now with a positive slope) and that there is a interest rate that is so high that only the most robust firm could afford a loan.
This relation is given by Eq. \ref{eq:Nponzi of i}: 
\begin{equation}
N_{ponzi}(t)=\left( \frac{i(t)}{i_{max}} \right) ^ \beta \times N_{tot}(t)
\label{eq:Nponzi of i}
\end{equation}
where:
	\begin{itemize}
	\item $i_{max}$ is the maximum resilience, a constant that reflects the interest rate that the most robust speculative company could stand without becoming ponzi;
	\item $\beta$  is the heterogeneity coefficient of firms 'ponziness', a positive number, which is the slope of the cumulative probability (power law) distribution of the resilience of the firms in the system that a firm will have resilience smaller than a specific value, and thus become ponzi with an interest rate at this specific value.
	\end{itemize}

In the first regime it is assumed that banks change the interest rate charged for loans if the demand for credit (expressed through the number of hedge firms) has changed, providing the feedback mechanism. In the second regime, the banks increase the interest rate if there is an increase in the number of their ponzi clients that are not fulfilling their debt obligations. So, an increase in the proportion of ponzi firms will lead to an increase the interest rate. This positive feedback loop is the opposite of the stable equilibrium obtained in the first regime.

In the original model with a network of firms \cite{Solomon 2013}, only once ponzi firms are recognized as ponzi by banks, they face an interest rate increase. In Section \ref{sec:Empirical}, banks did not recognize the fragility – low resiliency - of their clients in 2006, just before the subprime crisis. In our data, interest rates are rising due to central bank tightening. 

The mechanism that the banks apply when adjusting their interest rate may be modelled in different ways.  In our current analysis we will use a specific power law relation between the number of ponzi firms at a  time $t$ and the interest rate $i_{t+1}$  that they induce at the next time period. In the first regime we assume the interest rate is a function of the credit demand/supply expressed by the volume of loans:
the higher the volume of loans, the higher the interest rate in the next period. This relationship we again assume to be a power law 
so that a log plot of the volume of loans against the interest rate 
has a slope $\alpha$:
\begin{equation}
i_{t+1}=i_{min} \times (N_{loans}(t)/N_{tot}(t))^\alpha,
\label{eq:i of Nloans}
\end{equation}
where $\alpha$ can be either  positive or   negative, depending on whether the financial institutions  work with decreasing or with increasing returns to scale. In   recent times, it has become  common for them to have increasing returns to scale so the higher the demand for credit, the lower is the interest rate; this in turn makes the demand even higher, etc. This feedback loop can work in one of two modes: convergent (when $|\alpha \mu|<1$) or divergent $(|\alpha \mu|>1)$, which is discussed in \cite{Solomon 2013}. The latter case has been labelled as an `irrational exuberance' phase \cite{Shiller 2005}, which, according to the theory, is the state preceding the Minsky moment.

In the crisis regime, the interest rate is a function of the number of ponzi firms:
\begin{equation}
i_{t+1}=i_{max} \times (N_{ponzi}(t)/N_{tot}(t))^\alpha
\label{eq:i of Nponzi}
\end{equation}
where $\alpha$ is the inter-scale coefficient, which expresses the strength of the autocatalytic loop between the number of ponzi firms and interest rate changes. In our crisis accelerator model, we assume that when the number of risky firms increases, the interest rate increases too, to compensate the banks for the added risk of firms not being able to pay even the interest on their loans. Thus $\alpha$ is a positive number. 

In the crisis regime, the two coupled mechanisms,  Eqs. \ref{eq:Nponzi of i} and Eqs. \ref{eq:i of Nponzi}, generate a dynamic iterative process which,  as in the case of Walras, may either converge towards an equilibrium in both price, $i_{fixed}$,  and  level of ponzi firms, $N_{fixed}$; or, alternatively, may, after an exogenous shock, such as  a one-time injection of an increase in the proportion of ponzi  firms in the system,  develop into a divergent dynamical process, in which the number of ponzis and the interest rate both increase until the entire economy becomes ponzi.   

The count of the number of ponzi, risky firms, in  equation \ref{eq:i of Nponzi}, should be augmented by adding the number of firms that have failed, since such companies most probably did not fulfil their debt obligations towards their banks who, together with the firm's   trading partners, are clearly aware of their failure;  this point  was not  discussed in the original model \cite{Solomon 2013}.  

In the model, firms  are embedded in an inter-firm network, so the percolation of a crisis can be modelled. A financially fragile  firm is in  jeopardy; a  ponzi firm may fail by contagion as soon as one of its trade-connected partners fails. Using this rule, one can predict how failures propagate across the network. As the theoretical postulates of this model are not a subject of this paper, we redirect the reader interested in the concept of percolation to the paper \cite{Solomon 2013} and the references there in. Here we just quickly remind of the model equation.

The generic formula for the spread through contagion of any feature, e.g. an illness but here financial failure of a firm, valid for a very large class of networks (including all regular lattices, random Erd\H{o}s-R\'{e}nyi networks with any average number of neighbours, small-world networks, etc.) which have a finite average number of neighbours per node,  depends on only two parameters of the network: its critical density, $\rho_C$, and the peer-to-peer parameter (network exponent) $\gamma$, as:
\begin{equation}
N_{failed}=S \left[ 1- \frac{\rho(t)}{\rho_C} \right] ^ {-\gamma},
\label{eq:N of rho}
\end{equation}
where S>0 is a constant related with the exact way the contagion process is initialized. 
Note that as the density (here of ponzi firms), $\rho(t)$, approaches from below a critical value $\rho_C$,
the number of failed firms increases. When the density reaches the critical value, the system explodes, here every firm becomes ponzi and fails.
Recall that in general ponzi firms have not (yet) failed; firms that are ponzi are prone to failure, and do so through  contagion from a supplier or buyer who fails.
The actual number of failed firms at the next time interval depends on the ponzi density  at time $t$, $\rho(t)$.
The ponzi density at a time $t$ is, by Eq. \ref{eq:Nponzi of i}:
\begin{equation}
\rho(t)=\frac{N_{ponzi}(t)}{N_{tot}(t)}=\left( \frac{i(t)}{i_{max}} \right)^ \beta.
\label{eq:density}
\end{equation}
Similarly, the critical density is, also by Eq. \ref{eq:Nponzi of i},  a function of the maximum resilience in the data set and the critical interest rate $i_C$ for which the crisis propagates across the entire network, as thus the entire system has `percolated' into the state of a crisis. 
\begin{equation}
\rho_C=\frac{N_{ponziC}}{N_{tot}(t)}=\left( \frac{i_C(t)}{i_{max}} \right)^ \beta.
\label{eq:density}
\end{equation}
Around the critical density, $\rho_C$, the system displays extreme fluctuations in the number of failed firms. Below the critical density, the number of failed firms is small compared to the number of ponzi firms 

\section{Data set}
\label{sec:Data}
The basic quantities defined in Section \ref{sec:AMF model} are available as such in firms' balance sheets. It is much more challenging to create the links in the network of debtors and lenders.  In general, they have to be computed or evaluated by proxies from  available data. Banks play a pivotal role in the financial side of the economy, in particular in allocating and intermediating financial resources. Thus it is natural that our data on firms has been sourced from the banking system. 

The data bases we use are the property of one of the main Italian banks. The bank has requested not to be named in the paper. The analyses reported in this paper have been performed in the bank in a fully anonymous way, fulfilling both the data license policy, the privacy constraints and the bank's research department policy. Only aggregated data has been disseminated and distributed to members of the research group outside the bank.

The databases of the banking system are structured in a way dictated by  daily banking practice. Thus we have had to adapt them to our needs as described in the present section. In particular we have had to merge the data on the  balance sheets of individual firms  with a different database that documents the trade credit and payments between firms. The merger takes the form of a network structure whose  nodes  are characterized by the balance sheet database while the links are characterized by the trade credit database.

\begin{paragraph}{Balance sheet database (DB)}
is a proprietary database of the end-of-year balance sheet and Profit \& Loss statements of Italian limited liability companies for the period 2002-2009.
\end{paragraph}

\begin{paragraph}{The trade credit database (TC)}
 includes  goods/services supplied by a supplier to a buyer, as follows (without double counting):

\begin{itemize}
\item Trade invoices (accounts receivable) issued by a supplier $i$,  a client of the bank,  to a buyer $j$  that have been discounted (purchased by the bank from $i$ for less than their face value) in 2007. This is called a `non-recourse loan' since the lender (the bank) not only assumes title to the invoice, but also assumes most of the default risk since,  if there is a  default, the bank does not have recourse against the supplier $i$ but  must seek recourse with the buyer  $j$. Discounting-without-recourse has in effect rewired the expected future cash flow from being  $j$ to $i$, to a payment from $j$ to the bank itself. The discount rate will be a function of the Probability of Default (PD) of firm $j$.
\item Trade invoices (accounts receivable) issued by supplier $i$,  a client of the bank, to a buyer $j$ that are have been discounted and used as collateral for a loan, with recourse, from the bank to firm $i$, in 2007. Discounting-with-recourse means that the bank will hold the supplier $i$ responsible for repayment of the loan even if the buyer $j$ defaults on the payment;  the bank advanced a loan to  firm $i$  that $i$ has to repay in the future. The obligation for the trade payment  remains from  $j$ and $i$. The discount rate charged by the bank will be a function of the Probability of Default (PD) of supplier $i$.
\item Trade credit payments at maturity made in 2007 by buyer $j$, being a client of the bank, directly to supplier $i$ for goods/services. We don't know whether such invoices have never had their invoices discounted, or they were discounted by another financial institution.
\end{itemize}
We  represent this data in a network of trade transactions. The network is directional. We chose the direction of the link to denote the direction of credit owed or expected future money flow, so the arrow goes from trade debtor to  trade creditor. 

In general, trade creditors do not charge interest unless payment is delayed well beyond the settlement date. By delaying payment to a trade creditor, a business holds onto its cash  for longer; trade creditors are seen (wrongly) as a `free' source of capital. Some firms habitually delay payment to creditors in order to enhance their cash balance -- perhaps a short sighted policy. 
The period allowed before the invoice should be settled will vary from industry to industry. In the building trade in Italy, it is common for trade creditors to require settlement of invoices after 30 days,   while in some industries suppliers extend the time taken  up to 90 days (3 months).
\end{paragraph} 

\begin{paragraph}{Selection of the firms for the network.}
Each of the firms that appear in the TC data set have an identification code that allows us to connect the transaction records to their Balance Sheet and Profit \& Loss statements. By combining the  TC and BS datasets we  obtain the information  needed for the estimating the parameters in our model. However,  the two data sets are not perfect; not all firms that are available in the TC set are available in the BC set, and vice-versa. Therefore, we have imposed a condition on the selection of nodes for our analysis:  the   suppliers in the TC database are only selected if the total of their invoices in the TC database is at least 50\% of the supplier's annual sales. This results in a much smaller network then the original network of the trade credit that is being mediated by the bank.
\end{paragraph}

\section{Implementation of the proposed model}
\label{sec:Empirical}
We have  classified firms in our  database using the BS database that is closely aligned with Minsky's classifications:  `hedge', `speculative' and `ponzi.'. 
 We consider a firm to be `hedge', i.e. one  whose operating profit is sufficient to repay its current loans, if:
\begin{equation}
EBIT(t) \geq BL(t),
\label{eq:hedge}
\end{equation}
where: 
\begin{itemize}
\item[-] EBIT(t) is Earnings Before Interest and Taxes due in year t,
\item[-] BL(t) is Bank Loans, i.e. loans (debts to banks) incurred in year t. 
\end{itemize}

Speculative firms are those that are  able to cover the costs of their capital from  their gross operating profit before allowing for depreciation of assets; they need to run down their Current Assets or roll over at least a portion of their bank loans that fall due in the year in order to cover these costs. The continuing functioning of such firms depends on their ability to roll over liabilities. We classified firms as being speculative if:
\begin{equation}
(EBIT(t) < BL(t)) \wedge
(EBTDA(t) \geq FC(t)), 
\label{eq:speculative}
\end{equation}
where EBTDA(t) is Earnings Before Interest, Taxes due and Depreciation Allowance in year t.

For ponzi firms, ``the current income portion of payment commitments exceeds the current income portion of cash receipts", i.e. even if there is a gross operating profit, before deduction of the Depreciation Allowance, it was insufficient to cover Financial Costs \cite{EBTDA}:
\begin{equation}
(EBIT(t) < BL(t) ) \wedge
(EBTDA(t) < FC(t)). 
\label{eq:ponzi}
\end{equation}

The status of `speculative'  is very sensitive to changes in the  interest rate; such firms can easily  can easily become ponzi if interest rate rises. For ponzi and many speculative firms, working capital is negative, and the firm has to borrow to continue its operation. Borrowing will be tilted highly to short term loans, and profits are not enough to pay for interest rate and principle payments. The firm is highly vulnerable to a cash flow crisis.

\subsection{Empirical realization of the basic (non-network) model -- estimation of $\alpha$, $\mu$ and $\beta$ coefficients}
\label{sec:No network}
In this section we will couple the non-network model introduced in Section \ref{sec:AMF model} with the available data set. We need to connect  the iterative processes given by Eqs. \ref{eq:Nponzi of i} and \ref{eq:i of Nponzi}, or  Eqs. \ref{eq:Nloans of i} and \ref{eq:i of Nloans} with the data. 

\begin{figure}
\centering
  \includegraphics[scale=0.7]{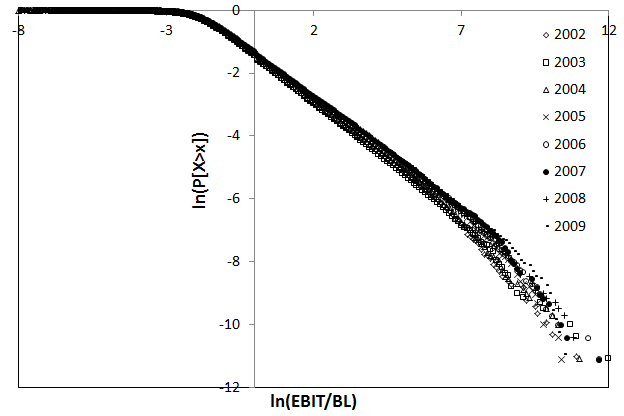}
\caption{The cumulative distributions $P[\ln(EBIT/BL)>x]$ on the $y$ axis against the values of $\ln(EBIT/BL)$ on the $x$ axis. For the calculation of $\mu$, we ignore the values $\ln(EBIT/BL)<-1.5$. From the slopes in the part of the graph in the range $\ln(EBIT/BL)>-1.5$ we determine the $\mu$ and $i_{min}$ coefficients as reported in Table \ref{tab:mubeta} with the very high goodness of fit, $R^2>0.99$.}
\label{fig:Mu Data}      
\end{figure}

\paragraph{Estimation of $\mu$ and $i_{min}$.}{The non-network model in the first regime has  three parameters that have to be estimated: the minimal resilience $i_{min}$, the inter-scale parameter $\alpha$ and the resilience heterogeneity parameter $\mu$, a negative number, being the slope of a Pareto law distribution of the resilience of the firms in the system, a cumulative probability that a particular firm will have resilience larger than a specific value, and thus be capable to take a new loan with an interest rate at this specific value. Depending on their values, the Equations \ref{eq:Nloans of i} and \ref{eq:i of Nloans} will describe a process that may either be divergent, or convergent  towards an equilibrium interest rate $i_{fixed}$  and an equilibrium level of loans $N_{fixed}$. 

These parameters can be directly measured from the data, in the following way. We have firstly measured the $EBIT/BL$ ratio for all the firms in the data set, for all years of our interest. The data show that the fragility indicator $EBIT/BL$ follows a power law in the range $x>exp(-1.5)=0.22$. Plotting the cumulative distributions $P[X>x]$ on the $y$ axis against the values of $EBIT/BL$ on the $x$ axis on a log/log plot (Figure \ref{fig:Mu Data}), and fitting (OLS) the data with the linear function, we obtained the $\mu$ and $i_{min}$ values as reported in Table \ref{tab:mubeta}.  $i_{min}$ has been fitted to simulate the empirical interest rate value.}
\begin{table}
\caption{Estimation of the parameters $\mu$, $i_{min}$ from the distribution of $EBIT/FC$ and $\beta$, $i_{max}$ from the distribution of $EBTDA/FC$.}
\label{tab:mubeta}    
\begin{tabular}{lllll}
\hline\noalign{\smallskip}
Year & $\mu$ & $i_{min}$ & $\beta$ & $i_{max}$ \\
\noalign{\smallskip}\hline\noalign{\smallskip}
2002 & -0.83 & 2.42 & 1.29 & 49 \\ 
2003 & -0.79 & 2.42 & 1.29 & 49 \\ 
2004 & -0.79 & 2.42 & 1.30 & 49 \\ 
2005 & -0.77 & 2.42 & 1.32 & 49 \\ 
2006 & -0.76 & 2.42 & 1.30 & 49 \\ 
2007 & -0.76 & 2.42 & 1.28 & 49 \\ 
2008 & -0.76 & 2.42 & 1.27 & 49 \\ 
2009 & -0.73 & 2.42 & 1.27 & 49 \\
\noalign{\smallskip}\hline
\end{tabular}
\end{table}

\begin{figure}
\centering
  \includegraphics[scale=0.7]{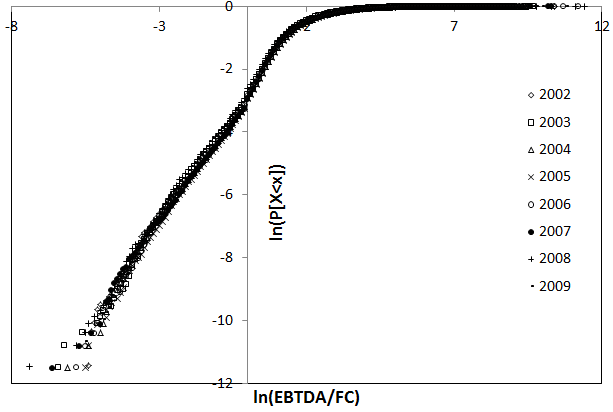}
\caption{The cumulative distributions $P[\ln(EBTDA/FC)<x]$ on the $y$ axis against the values of $\ln(EBTDA/FC)$ on the $x$ axis. For the calculation of $\beta$, we ignore the values $\ln(EBIT/FC)>3$. From the slopes in the part of the graph in the range $\ln(EBIT/FC)<3$ we determine the $\beta$ and $i_{max}$ coefficients as reported in Table \ref{tab:mubeta} with the very high goodness of fit, $R^2 \geq 0.98$. The shape of the distribution is disturbed around the value of 1, which however did not significantly disturb the fit.}
\label{fig:Beta Data}      
\end{figure}

\paragraph{Estimation of $\beta$ and $i_{max}$.}{The non-network model in the second regime  has three parameters that have to be estimated: the maximal resilience $i_{max}$, the inter-scale parameter $\alpha$ and the ponziness heterogeneity parameter $\beta$. $\beta$ is the slope of the cumulative Pareto law distribution of the resilience of the firms in the system, i.e. probability that a firm will have resilience smaller than a specified value, and as such it is a positive number. Depending on the values of the parameters, Eqs. \ref{eq:Nponzi of i} and \ref{eq:i of Nponzi} or Eqs. \ref{eq:Nloans of i} and \ref{eq:i of Nloans} will describe a process that also may  either diverge or  converge towards an equilibrium interest rate $i_{fixed}$  and an equilibrium level of ponzi companies $N_{fixed}$. 

This parameter can also be directly measured from the data, with a procedure which is similar, but yet significantly different than the one for $\mu$ . To estimate $\beta$, we have measured the $EBTDA/FC$ ratio for all firms in the data set and all years between 2002 and 2009. In all cases, the fragility indicator $EBTDA/FC$ follows a power law in the range $x<exp(3)=20$. Plotting the cumulative distributions $P[X<x]$ on the $y$ axis against the values of $EBTDA/FC$ on the $x$ axis on a log/log plot (Figure \ref{fig:Beta Data}), and fitting (OLS) the data with the linear function, we obtained the $\beta$ and $i_{max}$ values as reported in Table \ref{tab:mubeta}. $i_{max}$ has been fitted to simulate the empirical interest rate value.}

\paragraph{Estimation of $\alpha$.}{In order to estimate parameter $\alpha$, we need first to validate that the interest rate is $i_t$, at a each moment $t$ within a specific period of time, is a function of the number of ponzi companies / loans, expressed in neutral terms as $N_{t}$, according either to Eqs. \ref{eq:i of Nponzi} or Eq. \ref{eq:i of Nloans}. 

So we first counted the total number of firms and the number of ponzi firms for each year in our data set. The results are shown in Table \ref{tab:Ponzi}. We have also calculated the ponzi density (last column) as the number of ponzi divided by the total number of firms. Note that Eq. \ref{eq:Nloans of i} can be easily rewritten in terms of loans density:
\begin{equation}
\frac{N_{loans}}{N_{tot}}=\left( \frac{i}{i_{min}} \right)^{\mu}.
\end{equation}
Similarly,  
Eq. \ref{eq:Nponzi of i} can be easily rewritten in terms of risky companies density:
\begin{equation}
\frac{N_{ponzi}}{N_{tot}}=\left( \frac{i}{i_{max}} \right)^{\beta}.
\end{equation}
\begin{table}
\caption{Total number of firms and the number of ponzi firms per year in our data set. The ponzi firms are defined by the Eq. \ref{eq:ponzi}.}
\label{tab:Ponzi}    
\begin{tabular}{lllllll}
\hline\noalign{\smallskip}
Year & Nr. firms & Growth Nr. & Nr. hedge & Hedge density & Nr. ponzi  & Ponzi density\\ 
\noalign{\smallskip}\hline\noalign{\smallskip}
2002 & 469893 & - & 232432 & 0.49 & 83665 & 0.18\\ 
2003 & 521743 & 1.11 &255522 & 0.49 & 90297 & 0.17\\ 
2004 & 550947 & 1.06  &279562 & 0.51 & 89034 & 0.16\\ 
2005 & 592331 &  1.08 &310031 & 0.52 & 94011 & 0.16\\ 
2006 & 603799 &  1.02&322592 & 0.53 & 91169 & 0.15\\ 
2007 & 601535 & 0.99&316822 & 0.53 & 94095 & 0.16\\ 
2008 & 589141 & 0.98 &284366 & 0.48 & 112498 & 0.19\\ 
2009 & 565311 & 0.96&256429 & 0.45 & 121613 & 0.22\\
\noalign{\smallskip}\hline
\end{tabular}
\end{table}

The second unknown parameter in Eqs. \ref{eq:i of Nponzi} and \ref{eq:i of Nloans} is the interest rate. There are different interest rates that could possibly be relevant for the model, such as REFI, the policy rate, or Eurbor, the inter-bank loans rate. We have however decided to test our model against the interest rates for the loans to businesses, as reported by the bank of Italy for each year, table TTI30300, column "Bank interest rate on loans to business". The values are provided in Figures \ref{fig:Firm level}, \ref{fig:Hedge data} and \ref{fig:Ponzi data} and in the reference \cite{BankofItaly}. Please note that this interest is significantly higher then the inter-bank loans interest rate and that choosing the inter-bank rate would produce different model parameters. The reason we chose for the interest rate interest rate on loans to businesses is that it is closest to what the most of the (rather small) firms in our data set dealt with.

\begin{figure}
\centering
  \includegraphics[scale=0.6]{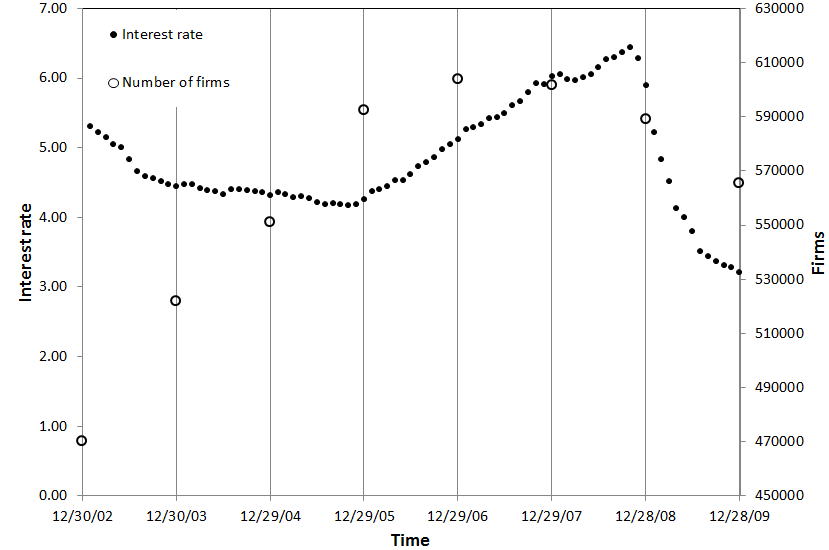}
\caption{Data that we use to estimate the model parameters. The interest rate (dashed line) comes from the Bank of Italy, table TTI30300, column "Bank interest rate on loans to business" \cite{BankofItaly}. The yearly data on the total number of firms, as reported each year on 31.12. is given by the white circles $(N_{tot})$. The total number of firms flattens already between 2005-2006 i.e. exactly when (and possibly because of) the interest rate starts increasing.
This is consistent both with our model, and with the ISML model \cite{Hicks 1980} \cite{Mankiw 2006}: 1.) More interest rate means the number of loans (for starting companies) diminishes. So one has less new companies. 2.) More interest rate means the number of failures (for previously marginal or ponzi companies) increases: more companies  disappearing from the list. 3.) In ISML-like interpretation: more interest rate makes less projects profitable enough to justify / support the interest payments. This means  less loans taken , less companies established, more companies closed. The fact that after 2008 the number of companies continues to decrease in spite of the decrease in interest rate is because
they have already loans that they cannot pay because the business is bad. So it is no point to take more loans.
Just to close the business (or not to start a new one).}
\label{fig:Firm level}      
\end{figure}
\begin{figure}
\centering
  \includegraphics[scale=0.6]{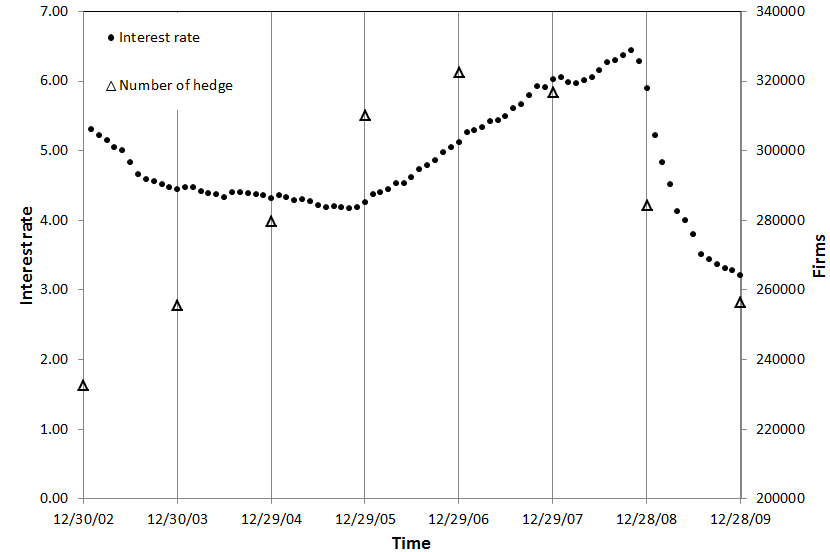}
\caption{Data that we use to estimate the model parameters. The interest rate (dashed line) comes from the Bank of Italy, table TTI30300, column "Bank interest rate on loans to business". The yearly data on the number of hedge firms, is calculated according to Eq. \ref{eq:hedge} and represented by the white triangles. The hedge firms start going down only from 2006, and not 2005 like the total number of firms of the previous Figure \ref{fig:Firm level}.
This means that first in 2005 people slowed down starting new companies (according previous graph), because they saw interest rate going up.
But the old companies started to disappear or to loose their hedge status 'only' from 2006 on,
 when the increase in interest rate started to actual hit them rather than just threatening their future.
This hypothesis can be in principle checked: whether there are less new companies established in 2005-2006 
(but not many companies loosing hedge status in this period).  And only in 2006-2007 starts loosing of status of
already existing companies.}
\label{fig:Hedge data}      
\end{figure}
\begin{figure}
\centering
  \includegraphics[scale=0.6]{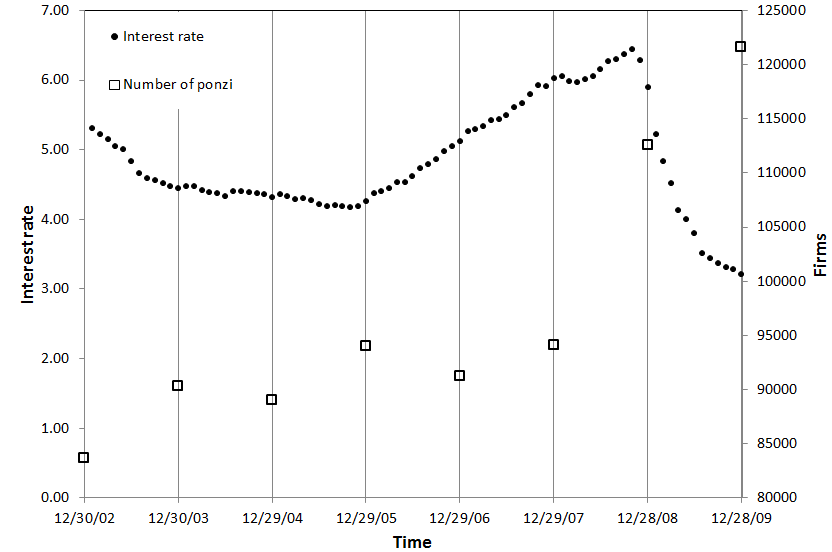}
\caption{Data that we use to estimate the $\alpha$ parameter. The interest rate (dashed line) comes from the Bank of Italy, table TTI30300, column "Bank interest rate on loans to business". The yearly data on the number of ponzi firms, has been calculated by according to Eq. \ref{eq:ponzi} and is given by the white squares. This graph completes the narrative of the Figures \ref{fig:Firm level} and \ref{fig:Hedge data}: while the year 2005-2006 was the year of new (starting) companies, and the 2006-2007 was the year of companies loosing their hedge (very solid) status, the year 2007-2008 was the year where former hedge and speculative companies became ponzi. 
When this happened, and ONLY when this happened the (central) banks
lowered exogenously the interest rate. 
But this was too late: the real sector had already problems of credit 
and of demand and the avalanche of ponzi continued (probably by contagion / percolation).}
\label{fig:Ponzi data}      
\end{figure}

From the monthly interest rate data, we see that the system passes through different phases in this relatively short period. Comparing the number of firms in the database as presented in Figure \ref{fig:Firm level}  and the monthly interest rate we can observe that:
\begin{itemize}
\item the system had reached a stable phase in 2003 and kept it until the end 2005, with the interest rate falling slowly but the number of firms increasing. This behaviour is typical for the \emph{Loan accelerator} model, when the negative coefficient $\mu$ in Eq. \ref{eq:Nloans of i} is coupled with the negative coefficient $\alpha$ in Eq. \ref{eq:i of Nloans}. 
\item In 2006 a transition happened which is not difficult to understand: we can observe an exogenous change in the policy rate,  which could lead to an increase in the probability of being ponzi (we think that this was also accompanied by a reduction in the risk premium around January 2006). The banks expected that the system was in a good economic phase of the cycle (meaning that they expected that  firms' EBTDA would increase), and that the firms would be able to repay their financial costs, thus reducing the probability of being ponzi (which is  what we measure).  However, we can also observe that the total number of firms starts to decrease. Thus, the interest rate rose, but both the total number and the proportion of ponzi firms fell. This is a beginning of the Minsky crisis: the firms change their attitude and demand less credit.
\item In 2007 and 2008, the system is  in a state of a financial fragility as defined by the Minsky \emph{Crisis accelerator} model, when the interest rate (positive $\alpha$) and the number of ponzi (positive $\beta$) both grow, while the total number of firms drops. 
\item In 2009 the interest rate dropped dramatically, due to an well-known exogenous reason. This however did not induce an immediate recovery, and it is our assumption that this is due to the network effects. Namely, the total number of firms in the system started shrinking in an accelerated way when the ponzi density became sufficiently high. The large number of ponzi firms arising from the credit crunch crisis in 2008, started failing and leaving an equally large number of remaining firms unpaid, and because of this, the number of ponzi firms further increased in the year 2009.
\end{itemize}
 }

\paragraph{Simulation results.} {We have fitted the measured number of firms and the given interest rate (Figure \ref{fig:Interest Rate Sim}), according to the different models as introduced in Figure \ref{fig:Firm level}.

We have further applied both the `Loans accelerator' and `Crisis accelerator' models. The heterogeneity parameters as shown in Tables \ref{tab:mubeta} and \ref{tab:results} have been used. The value of the heterogeneity coefficient $\alpha$ has been optimized to satisfy the empirical interest values. $\alpha$ has been re-calculated in January each year and kept constant until the following January. The exact values of all parameters are given in Table \ref{tab:results}. 

In the period 2003-2005, the system is close to a fixed point, as given by either Eqs. \ref{eq:Nloans of i} and \ref{eq:i of Nloans}, or Eqs. \ref{eq:Nponzi of i} and \ref{eq:i of Nponzi}, the exponents are such that their product is very close to 1: $\alpha \beta \sim 1$ condition has been satisfied (2003=1.006, 2004=1.001; 2005=1.003). This means that the Crisis accelerator process has been convergent. 

However, this fixed point could not have been sustained  because of the change of the policy of interest rate at the beginning on 2006. The change in the interest rate launched a period of `irrational exuberance' in 2006.
In the period 2006-2007 the fast `increasing returns' process took place, i.e.  the  interest rate falls, while the demand (expressed as the proportion of firms taking loans) grows.  $\alpha \mu > 1$ condition is satisfied (2006=1.02, 2007=1.02).

\begin{table}
\caption{Simulation results, using the previously fitted parameters $\beta$, $\mu$, $i_{max}$, $i_{min}$ and the models of Loans accelerator, Eqs. \ref{eq:Nloans of i} and \ref{eq:i of Nloans} and Crisis accelerator, Eqs. \ref{eq:Nponzi of i} and \ref{eq:i of Nponzi}. The results are: the estimation of the parameter $\alpha$ for each year, starting from January until December, leading to the $N_{loans}/N_{tot}$ and $N_{ponzi}/N_{tot}$ ratios in December of each given year. This was necessary because of the yearly basis on which the coefficients $\beta$ and $\mu$ are measured.}
\label{tab:results}    
\begin{tabular}{lllllll}
\hline\noalign{\smallskip}
Year & $\alpha_1$ & $N_{loans}/N_{tot}$ & $\alpha_1 \mu$ & $\alpha_2$ & $N_{ponzi}/N_{tot}$ & $\alpha_2 \beta$\\ 
\noalign{\smallskip}\hline\noalign{\smallskip}
2003 & - & 0.53  & - & - & 0.18 & - \\ 
2003 & -1.235 & 0.62 & 0.976 & 0.78 & 0.15 & 1.006 \\ 
2004 & -1.262 & 0.63 & 0.997 & 0.77 & 0.15 & 1.001\\ 
2005 & -1.293 & 0.65 & 0.995 & 0.76 & 0.16 & 1.003\\ 
2006 & -1.346 & 0.57 & 1.023& 0.765 & 0.18 & 0.994\\ 
2007 & -1.338 & 0.50 &1.017 & 0.775 & 0.20& 0.992\\ 
2008 & -1.325 & 0.47 & 1.007& 0.85 & 0.20 & 0.997\\ 
2009 & -1.242 &0.81 & 0.907 & 0.795 & 0.11 & 1.009\\
\noalign{\smallskip}\hline
\end{tabular}
\end{table}



Due to the network effects, a contagion of failures started to take place in 2008 and the number  of companies started to drop.

Finally, in 2009, thanks to another policy intervention of the European Central Bank, the regulation took place and the interest rate dropped exogenously by a very large amount. This however could not bring the desired healing effect immediately, for the reason that the failure contagion process that started in 2008 could not have been inverted: the failed companies have left their debtors in a difficult situation which even the extremely low interest rate could not heal. The severely reduced number of firms, had long term consequences, possibly due to the smaller market, fewer customers etc. 
In this period we have applied the model of slow increasing returns $\mu \alpha <1$. The interest rate falls and the number of loans increases due to an convergent process. 

\begin{figure}
\centering
  \includegraphics[scale=0.7]{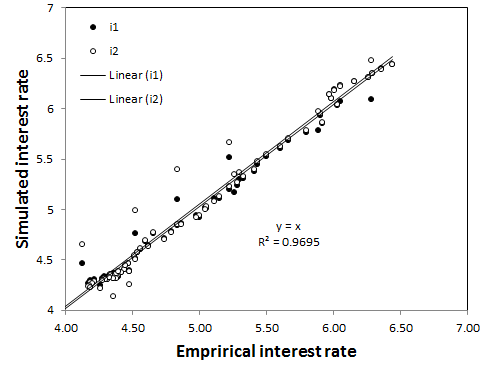}
\caption{On the x-axis is the interest rate as copied from the Bank of Italy database, table TTI30300, column "Bank interest rates on euro loans to businesses". On the y-axis is the simulated interest rate, $i_1$ for the regime 1 and $i_2$ for the regime 2. To show the goodness of simulation, we have fitted the points with the y=x line. The goodness of fit is $R^2=0.97$ which shows that the interest rate has been successfully simulated in both regimes.}
\label{fig:Interest Rate Sim}      
\end{figure}
}

\paragraph{Discussion}{The period of 2003/5 was the upside of the cycle (rational exuberance?) and 2008-09 the down side. The interest rates moved accordingly. The key question to answer is what slowed down the growth of economy, reflected in slowed increase in the number of companies in 2006? For what reason did the ECB increase the interest rate in the beginning of 2006 and take the system 
from an upside to the Minsky financial accelerator regime in 2007-9?
This happened long before the defaults on sub-prime mortgages reached astronomical proportions. We do not have an answer to that. 
}

\subsection{Empirical evidence of the network relevance and its current limitations }
\label{sec:Network}
In this Section we analyse the contagion of financial distress (not only of formal, legal failure or default, but even  ponzi status contagion) from customers (clients, buyers) on suppliers in a trade credit inter-firm network. We focus our analysis on the supplier firms belonging to the Manufacturing industrial classification. There are several reasons for this. One  is that both Constructions and Wholesale trade are at the `dead end' of the supply chain -- the correlation of their Purchases and Sales is different from that in the Manufacturing sector, which is the very heart of the economy. Also, since the amount of net sales for retail firms and other sectors are different in scale they cannot be directly compared to Manufacturing. This is because of the different business processes involved. In addition in our database  the number of selected suppliers from other sectors is rather small compared to the Manufacturing sector. 

\begin{figure}
\centering
  \includegraphics[scale=0.75]{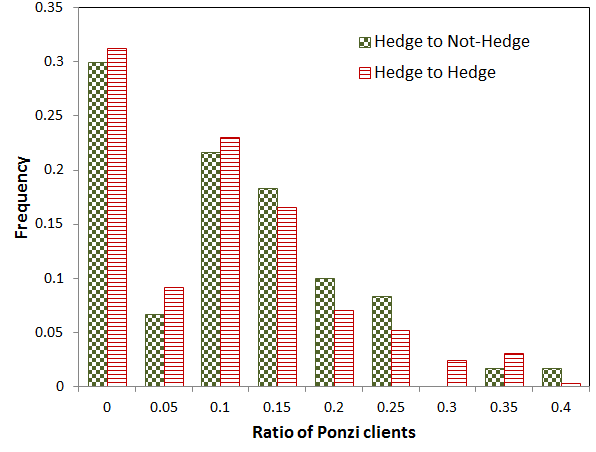}
\caption{Histograms of the ponzi buyers ratios for the two subgroups of companies: the hedge companies that remain hedge in the next year (red, stripe pattern) and the hedge companies that do not remain hedge in the next year (green, square pattern).}
\label{fig:PonziBuyers}      
\end{figure}

Until this point, the dynamics of the interest rate, loans and non-hedge companies has been found to be consistent with the Minsky dynamics without network obstructions:
each company influences the interest rate which in turn influences the individual companies irrespective of their clients status. 
As the crisis deepens, and companies start loosing business by their clients becoming  non-hedge one starts to see in addition to the interest rate intermediate feedback loop, direct influence of the status of the company trade partners. 
In particular, one sees in Figure \ref{fig:PonziBuyers} that there is a clear effect on the fraction of ponzi clients on the status of a company: 
for a fraction of ponzi clients less then 0.15 one sees that the number of companies that keep the hedge status dominates. 
On the contrary among companies with more then 0.15 fraction of ponzi clients, the companies that are downgraded to non-hedge status dominates. 
This explains the fact that by the time the density of ponzi exceeds 0.15 one has a strong acceleration in the rate of companies loosing the hedge status. 
In particular one sees in  Figure \ref{fig:Firm level} a great decrease after 2007 in the growth of the number of companies: 
from a sustained increase by 30,000 a year in the preceding period to a decrease of order of 20,000 a year after the Minsky moment of 2007.
This corresponds to the moment when the ponzi density suddenly exceeded 0.18: 0.18 for 2008 and 0.22 for 2009 according to Table \ref{tab:Ponzi}.
Combined with  Figure \ref{fig:PonziBuyers}, these numbers indicate that  direct contagion from one's clients starts to operate only by the time a company has $0.15*35.5 \approx 5$ ponzi neighbors.
This means that most likely the network effects are described by a bootstrap percolation process \cite{Kindler 2012} rather than by simple percolation \cite{Cantono 2010}, \cite{Solomon 2013}.

\begin{figure}
\centering
  \includegraphics[scale=0.6]{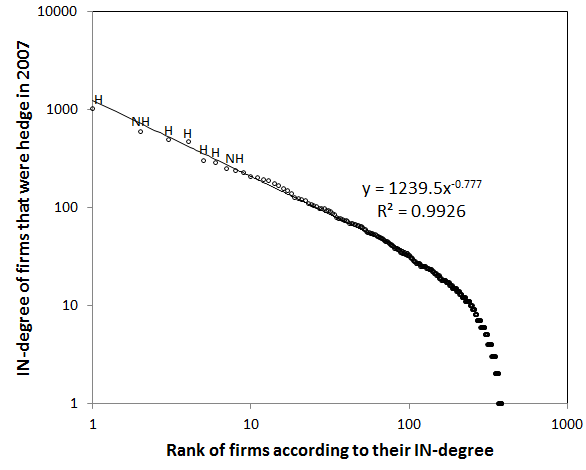}
\caption{IN degree of the selected supplier nodes, which have hedge status in 2007.
 Some of them change this status in the year 2008. For the companies with the highest IN-degree the status in 2008 is labeled: H for the ones which remain hedge and NH for the ones which change the status in 2008. The upper tail of this Zipf plot is fitted by the power law function given in the plot. 
The exponent of this Zipf plot of 0.77 corresponds to the Pareto exponent of 1.3 (by Pareto function we assume the Cumulative Probability Function $P[INdegree \geq x]$).  The largest IN-degree in this set is in the order of 1000. Average degree is 35.5.  }
\label{fig:INDegree}      
\end{figure}

The greatly reduced total number of firms in 2009 implies that many of the firms defaulted in that year -- even went bankrupt \footnote{Without an immediate cash flow from the "next generation", the suppliers could not afford to continue in business.}. The  contagion of defaults is the heart of the percolation model and we envisage to study in more detail the fate of the defaulted companies and of their partners. The current data can be interpreted as an indication that the defaults propagate rapidly when the density of ponzi is in the vicinity of the critical ponzi density $\rho_C = 0.18$. Indeed, from the data in Table \ref{tab:Ponzi} we see that this critical value was reached in 2008, when $\rho=0.19$. The number of defaulted firms in 2009 is very large, which is clear from the fact that the population of firms shrank by over 10\% in 2009. This number is almost as large as the total number of ponzi in 2008 (even though we absolutely do not imply here that all firms which defaulted in 2009 were ponzi in 2008).

It is very difficult with our partial information to characterize in detail the {\em significance of the defaults of ponzi firms on their neighbours}. Clearly, we don't have the data on defaults for the individual firms, neither do we have the bankruptcy data. The only indication of possible failure that we have is the `disappearance' of the firms from our database. In some cases this disappearance is also related with defaults, in many other cases it may be for other reasons. Therefore, the direct application of the model developed in \cite{Solomon 2013} and introduced in Section \ref{sec:AMF model}, which assumes the knowledge on the defaults, is not possible on the present data set. However, we propose here a method of that that might be able to give a measure of the strength of the network on the given data set. 

Our preliminary tests across the trade credit network showed that the status of individual companies switches from one year to another – from one year to another a ponzi firm can become hedge, and then go back again etc. The ponzi status is not at all exclusively reserved for the small companies. Sometimes, and for sure in the turbulent periods, large companies can also fail to fulfil the condition given by Eq. \ref{eq:hedge} and their interest on loans can become higher then their earnings. This empirical fact and data availability  led us to consider the contagion of `ponziness' rather then the contagion of failures. 

In order to measure the propagation of ponziness, we have used the sub-network formed of the nodes for which we know at least 50\% of their clients. The procedure for the selection of such nodes has been explained in Section \ref{sec:Data}. In order to measure growth of Sales/Purchases we need to perform 50\%-120\% matching for both years 2007 and 2008, which we have done.  From the initial selection of the suppliers for which we know more than 50\% of the buyers (1029 firms), approximately one quarter of the firms disappeared in the year 2008 from our database. These firms are not a subject of our analysis, but we simple want to give information about their number. This number is obviously very high, even for the year of a crisis such was the year 2007. 

Of the remaining 782 firms, a great part has been disqualified from the selection for our analysis as we have decided to discard all industrial sectors but Manufacturing. The final set of suppliers for which we know the majority of the buyers in 2007, and for which this ratio remains >50\% also in 2008, and in which the suppliers are only from the Manufacturing sector, has 462 firms. 

Applying the same procedure as in the previous section, we have performed the Minsky classification of suppliers. The large majority of these suppliers has hedge status in 2007: the number of hedge firms is 387. The number of speculative and ponzi firms in 2007 is significantly smaller: 32 and 43, respectively.

Because of our uncertainty about defaulting or disappearance of firms, we are only interested in the contagion of `ponziness' from buyers to suppliers. Our data do not allow us to measure the network effect as originally proposed in the model \cite{Solomon 2013}, for the reason that we do not have information about defaults. Nevertheless, we have managed to design and perform a number of other tests that can, in an implicit way, show the effect of ponzi firms in the network.

Our assumption is that a supplier that has a larger number of ponzi buyers has an increased chance of becoming ponzi itself, in case of the crisis, in particular a credit crunch crisis. This assumption is based on our our expectation that some of these ponzi buyers can easily default. To test it, we performed the following check:  for each of the selected supplier nodes which were hedge in 2007, we count the number of buyers that we know ($N_j$) and the number of them which are ponzi ($N_{ponzi}$). Then we can determine the ratio of ponzi buyers $N_{ponzi}/N_j$. Finally, we plot histograms of this ratio for the two sets: the hedge companies that stay hedge in the next year, and the hedge companies that do not stay hedge in the next year. These histograms are presented in Figure \ref{fig:PonziBuyers}. 
The histograms in Figure \ref{fig:PonziBuyers} show two things:
\begin{itemize}
\item The characteristic shape of the histograms shows a high peak at 0. This is due to the distribution of the IN-degree (Figure \ref{fig:INDegree}) which shows that the largest number of companies in our sample (but also in other samples that we have tested) has a very low IN-degree of 1 or 2. In combination with the fact that the percentage of ponzi companies in the sample is very low $(\sim 10\%)$,  means that there is a very high probability for a firm to have both a  small number of buyers and also a small number of ponzi buyers. This combination  justifies the size of the highest peak in the histograms which indicates that about 30\% of the companies in the set have zero ponzi buyers.
\item   Almost all other companies have ponzi buyers ratio between zero and 0.3. The rest are some rather incidental occurrences. 
\item The two histograms 'cross' between 0.10 and 0.15. Left of this crossing, the companies that remain hedge have higher frequency for each of the ponzi ratios (0, 0.05, 0.10). Right of that, the companies that will lose their hedge status have higher frequency at each of the ratios (0.15, 0.20, 0.25).
\end{itemize}

These observations show  in a qualitative way what is the impact of ponzi buyers/debtors to a supplier/creditor. Clearly, whatever the ratio of ponzi buyers a firm has, it has certain probability to become ponzi as well. This chance is greater when the ratio of ponzi buyers is larger. But, the differences are not very large and to express them in a quantitative way is very difficult. However, the increased chance of a firm becoming ponzi in the case when its proportion of ponzi buyers is larger, represents evidence that the network effect is present.  

To get further insight into the differences between these two groups of companies, we have compared the growth of the purchases of all the buyers of a supplier $i$, and in this way we have estimated the growth of sales of the supplier:
\begin{equation}
\mbox{Estimated Growth}_i=\sum_j{TC_{ji}*\frac{Purchases_j(2008)}{Purchases_j(2007)}}
\label{eq:sumPji}
\end{equation}
where $TC_{ji}$ is the weight of the network link between a buyer $j$ and a supplier $i$, equal to the sum of all trade credit transactions between them in 2007, with the growth of sales of suppliers as measured from their own balance sheets:
\begin{equation}
\mbox{Growth}_i=\frac{Sales_i(2008)}{Sales_i(2007)}
\label{eq:GS}
\end{equation}

\begin{figure}
 \centering
  \includegraphics[scale=0.55]{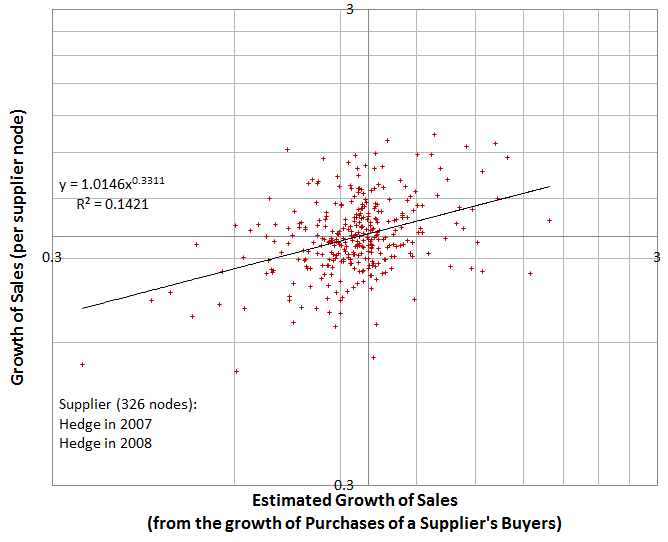}
\caption{Scatter plot of the Growth of Suppliers against their Estimated Growth for the set of firms which were hedge both in 2007 and in 2008.}
\label{fig:HedgeToHedge}      
\end{figure}

\begin{figure}
\centering
  \includegraphics[scale=0.55]{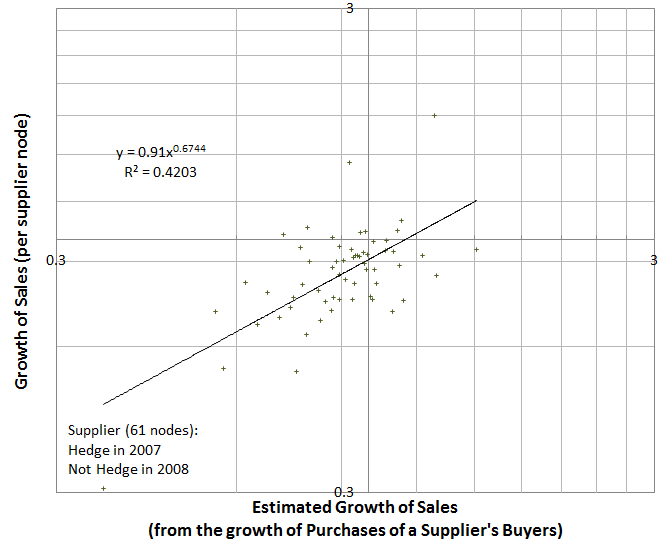}
\caption{Scatter plot of the Growth of Suppliers against their Estimated Growth for the set of firms which were hedge in 2007 but not in 2008.}
\label{fig:HedgeToNotHedge}      
\end{figure}

We expect to see strong correlations of the estimated growth and the sales growth as measured only in suppliers for which the network remains (largely) the same and the connections between buyers and suppliers are not broken. By performing such a test for these two groups (the hedge firms that remain hedge and the hedge firms that do not remain hedge) we have observed some differences which are shown in the scatter plots \ref{fig:HedgeToHedge} and \ref{fig:HedgeToNotHedge}.


These two scatter plots have been fitted with a power-law fitting curve for which the coefficients and the goodness of fit are given in the plots. The fits are not very good, and the residual yields a rather low $R^2$ value in Figure \ref{fig:HedgeToHedge}. However, the correlation is  higher for the suppliers which were hedge in 2007 but not in 2008; in Figure \ref{fig:HedgeToNotHedge}, $R^2$=0.42 tells us that over 40\% of the fluctuations in Sales of a supplier is caused by the fluctuations in the Purchases of its buyers.

 Unfortunately, we were not able to firmly validate the value of $\gamma$. Perhaps an additional evidence for it can be found in our working paper \cite{Golo2015}, where we report measurements of the contagion of growth from customers to a supplier, selecting firms according to their industrial classification, size or access to credit (rating), which produced $R^2$ values in the range 0.17-0.70. This led us to conclusion that the $R^2$ value of 0.15 (the smallest measured value) is the best possible indicator of $\gamma$.

\paragraph{Too dynamic to fail.} {An additional observation from Figure \ref{fig:HedgeToNotHedge} is that most of the suppliers, the hedge suppliers from 2007 that became not-hedge in 2008, had almost only negative growth and that it was mostly related with the negative growth of their buyers. The (green) dots in Figure \ref{fig:HedgeToNotHedge} are mostly concentrated in the third quadrant: the negative Estimated growth is accompanied with the negative Growth in sales. In contrast, the (red) dots in Figure \ref{fig:HedgeToHedge} are quite often found in the second quadrant too; the negative Estimated growth is accompanied with the positive Growth of Sales. This indicates that such a supplier has not been bound to only existing buyers in 2007, but it kept acquiring new purchasers in 2008 which compensated for the negative growth of the old ones.  

Then, we have also compared the growth correlation scatter plots for the other Minsky classes.  The correlation was similarly strong for the speculative suppliers; but for ponzi suppliers there is real zero correlation between the estimated growth and their own growth (of course these are only ponzi firms which survived). That also indicates that the ponzi firms that survive are the firms that find new buyers.

So the explanation for the measurements that we have presented lead us to the conclusion: too dynamic to fail.
The companies which show very low correlation on our data (yearly estimated growth against the real growth) are the companies that are finding their way out of the crisis; perhaps they are finding new trading partners.
The companies that are well-correlated are the companies which degraded their  status from Hedge to Not-Hedge. They are `unable' to find the new partners fast enough.

Thus the Minsky accelerator theory is confirmed in all its aspects by our empirical data.
As described in detail in \cite{Solomon 2013} once the empirical data confirm our theoretical model,
 the model can be used to make (limited and probabilistic) further prognosis on the expected evolution 
of the economy based on the current values of the economic parameters. 
Moreover the model may assist the work of the regulators to keep the economy from collapsing.
The key concept in this respect is the ponzi-percolation transition line.
This is a line in the plane whose axes are:
-  $rho_{Ponzi}$ the fraction of ponzi companies in the economy and
-  $i$, the interest rate.
Below that line the economy is stable while above it, the economy is in a Minsky accelerator state where both $rho_{Ponzi}$ and $i$ increase out of control. 
In terms of the micro-economic variables the two phases differ by the geometry of the clusters of connected / trading ponzi companies:
- in the stable phase each ponzi company is connected to hedge or speculative companies and only occasionally one or very few ponzi partners. 
- in the Minsky accelerator phase the ponzi companies form large connected clusters which can propagate fast eventual failure waves.
The problem faced by the regulator is :
1) watching that the economy does not leave the stability range (e.g. prevent the proliferation of the ponzi companies)
2) abstain from moves that can kick the system into the Minsky unstable phase.
These 2 objectives seems consistent and converging to similar policy recommendations.
However it turns out that they open a trap responsible for most of the documented crises [see for instance \cite{Kindleberger 2007} and \cite{Reinhart 2012}].

The picture in which the firms are organized in a static supply network on which failure contagion propagates is not entirely supported by the data. Only the companies that do not create new connections to new trade partners display failure contagion.
Thus one should think of the resilient part of the economy as an agent based model \cite{LLS 2000} in which the interactions between the agents are better likened to a hot gas of rapidly moving agents rather than a cold brittle crystal of frozen ice. It is the dynamic character of the trade connections between the firms that insures their survival. An economy characterized by a static trade network turns out to be too fragile to survive.
}

\section{Summary}
We have presented empirical evidence \cite{Solomon 2013} for an economic model that includes bottom-up, top-down and peer-to-peer feedback loops.
The interest rate is the “top” variable.  The “bottom” variables are the states of the firms.
The “top-down” interaction is the influence of the interest rate on the status of the individual firms:
an increase  in the interest rate can transform a hedge firm into a ponzi or even make a ponzi disappear altogether.
The “bottom-up” interaction is the influence of the individual states of the companies on the interest rate.
For instance a company moving from a hedge status to a ponzi status will contribute to the increase of the global interest rate.
The peer-to-peer interactions relate the change of status of a company to changes in in the status of its trade partners.
For instance a hedge company whose clients become ponzi may be brought into a non-hedge status if it does not initiate immediately new trade connections to new customers.

The model is heterogeneous, in the sense that each firm has a different resilience
level, according to a power law distribution. Moreover the network endows each of the agents with a different environment. In the sense of their generic behavior however,
the agents are homogeneous – with zero intelligence – as they can not learn
and they behave in the same way until the end of simulation. However we
have different simulation ‘modes’ (e.g. before and after the Minsky moment) 
in which we assign different behaviour to the agents/firms and the banks.

The general theoretical importance of the proposed mechanism is connected with the missing endogenous floor of FIH. Our approach, in line with Minsky's dynamic mood towards the representation of economic phenomena, underlines that the possibility to have an endogenous floor is due to the existence of a dynamic supply-network. This is obviously related to the nature of the our model and does not exclude other dynamic mechanisms leading to similar implications. 

Thus, unlike a very large number of standard models in the classical economic theory, our model has emergent properties. It also transcends the equilibrium framework in as far as it can simulate the run-away evolution in the unstable phase of the system. In such a phase the number of failed companies rapidly increases and doesn’t end in an equilibrium but rather in  a crisis. 

The model has been devised so as to be capable of including the inter-firm network of direct creditor-debtor relationships. This is the interaction that agents/firms can have upon one another. 
We used empirical data to create the network of trade credit and check the extent to which this is relevant for the economic dynamics and for crisis propagation. 
 We have compared our model to a data set, part of which is the property of a large bank. 

The data showed that the non-network part of the model is easily validated. 
 However the peer-to-peer network-effect which is supposed to intermediate the direct failure contagion between interacting firms takes place only at large ponzi densities and only for companies with static trade links.  As a consequence, the data could not establish a well defined quantification of the ‘peer to peer’ parameter because the network links are not the same at different times. We have developed the required procedure to detect and estimate the critical ponzi density should new systems with fixed geometry or new, more detailed, data become available. As it turns out, the present empirical system is both too dynamic to fail and also too dynamic for a static network picture.

\section*{Acknowledgements}
We thank the Institute for New Economic Thinking (INET), as this work has been performed with their support under grant ID INO1100017. We thank The Annual Workshop on Economic Science with Heterogeneous Interacting Agents (WEHIA) 2012 and 2013 organizers for comments on our preliminary results and for allowing us to present them here.

\end{document}